\documentclass[runningheads]{llncs}
\usepackage{graphicx}
\usepackage[backref]{hyperref} 
\hypersetup{hidelinks}
\usepackage{multirow} 

\begin{document}
\title{Exploring Vanilla U-Net for Lesion Segmentation from Whole-body FDG-PET/CT Scans}
%
%
\author{Jin Ye\inst{1} \and Haoyu Wang\inst{1,2} \and Ziyan Huang\inst{1,2} 
\and Zhongying Deng\inst{1,3} \and Yanzhou Su\inst{4} \and Can Tu\inst{1,2} 
\and Qian Wu\inst{1,5} 
\and Yuncheng Yang\inst{1,2} 
\and Meng Wei\inst{1}
\and Jingqi Niu\inst{1,2} \and Junjun He\inst{1,2}\thanks{Corresponding Author}}
\authorrunning{J. Ye et al.}
%
\institute{
1 Shanghai AI Lab, Shanghai, China \\
2 Shanghai Jiao Tong University, Shanghai, China \\
3 University of Surrey, Guildford GU2 7XH, United Kingdom \\
4 University of Electronic Science and Technology of China\\
5 College of Electronic and Information Engineering, Tongji University
}
\maketitle              
\begin{abstract}

Tumor lesion segmentation is one of the most important tasks in medical image analysis. 
In clinical practice, Fluorodeoxyglucose Pos-itron-Emission Tomography~(FDG-PET) is a widely used technique to identify and quantify metabolically active tumors. However, since FDG-PET scans only provide metabolic information, healthy tissue or benign disease with irregular glucose consumption may be mistaken for cancer. To handle this challenge, PET is commonly combined with Computed Tomography~(CT), with the CT used to obtain the anatomic structure of the patient. The combination of PET-based metabolic and CT-based anatomic information can contribute to better tumor segmentation results.
In this paper, we explore the potential of U-Net for lesion segmentation in whole-body FDG-PET/CT scans from three aspects, including network architecture, data preprocessing, and data augmentation.
The experimental results demonstrate that the vanilla U-Net with proper input shape can achieve satisfactory performance. Specifically, 
our method achieves first place in both preliminary and final leaderboards of the autoPET 2022 challenge. Our code is available at \url{https://github.com/Yejin0111/autoPET2022_Blackbean}.

\end{abstract}

\section{Introduction}
Fluorodeoxyglucose Positron-Emission Tomography~(FDG-PET) and Computed Tomography~(CT) are two vital modalities in the diagnostic workup for various malignant solid tumor entities. FDG-PET is widely used in oncology to reflect the glucose consumption of tissues for therapy control because tumor lesions typically have increased glucose consumption~\cite{2007The}. 
FDG-PET is a sensitive but non-tumor-specific method for identifying malignancy.
It can also be seen in healthy tissue or benign disease as inflammation or posttraumatic repair could be mistaken for cancer~\cite{2005False}. To overcome this problem, PET is commonly combined with CT to attach anatomic information to the metabolic data and improve diagnostic accuracy in the real clinical scenario. 
As a result, efficiently utilizing the paired PET and CT images can help accurately diagnose malignant tumors and avoid needless surgery. 

In this context, the Automatic Lesion Segmentation in Whole-Body FDG-PET/CT Challenge~(autoPET) is organized to offer the opportunity to participants to develop an automatic bi-modal approach for the three-dimensional segmentation of tumors lesions in FDG-PET and CT scans. The purpose of this challenge is to provide a large-scale training data set to promote the deep-learning-based automated tumor lesion segmentation on whole-body FDG-PET /CT data\footnote{https://autoPET.grand-challenge.org/}. The challenge provides more than 1000 cases of paired PET/CT scans, while the previously released PET/CT datasets usually contain less than 400 scans.

Deep learning methods with Convolutional Neural Networks (CNNs) have become \textit{de facto} standard in automated PET/CT lesion segmentation. However, the performance of tumor lesion segmentation in whole-body PET/CT remains far from satisfactory due to limited training data in previous works~\cite{2007The,2019DenseX}. Based on the large-scale training data provided in the autoPET Challenge, 
we provide some empirical analysis of the U-Net~\cite{10.1007/978-3-319-24574-4_28} for the lesion segmentation task, including model architecture, data pre-processing, and data augmentation.
The experimental results demonstrate that the vanilla U-Net with proper input shape can achieve satisfactory performance. Specifically, a Dice score of 92.81\%, on the preliminary test set.

\section{Material}
The challenge involves only one task, i.e., accurate and fast lesion segmentation. 

\subsection{Dataset}
The autoPET dataset consists of patients with histologically proven malignant melanoma, lymphoma, or lung cancer as well as negative control patients who were examined by FDG-PET/CT in two large medical centers (University Hospital Tübingen, Germany \& University Hospital of the LMU in Munich, Germany). All PET/CT data within this challenge have been acquired on advanced PET/CT scanners (Siemens Biograph mCT, mCT Flow and Biograph 64, GE Discovery 690) using standardized protocols following international guidelines. In detail, the dataset provides 1,014 paired FDG-PET and CT studies from 900 patients as the training data, and 200 studies as the testing data. Furthermore, there are 5 cases from the testing data that are provided for participants to evaluate their algorithms. 

\subsection{Metrics}
A combination of three metrics selected by the organizer to evaluate the methods: (1) Foreground Dice score of segmented lesions~(DSC); (2) Volume of false positive connected components that do not overlap with positives~(False Positive Volume, FPV); (3) Volume of positive connected components in the ground truth that do not overlap with the estimated segmentation mask~(False Negative Volume, FNV). The final ranking of submitted algorithms is computed by the numerical mean of the single rankings,
\begin{equation}
    \mathrm{score} = 0.5 \times \mathrm{Rank_{DSC}} + 0.25 \times \mathrm{Rank_{FPV}} + 0.25 \times \mathrm{Rank_{FNV}}.
\end{equation}

\section{Method}
\begin{figure}[t]
\centering
\includegraphics[width=\linewidth]{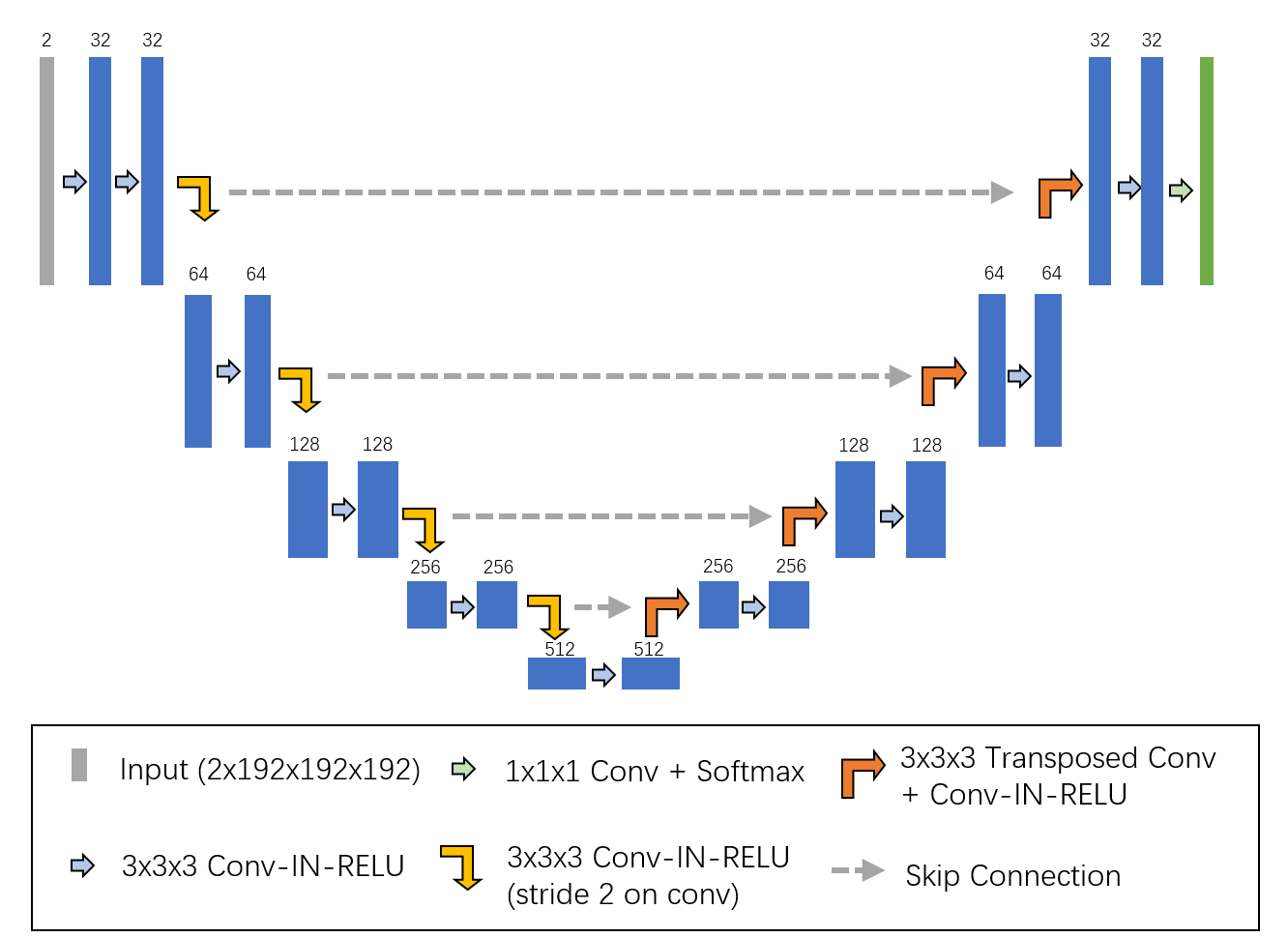}
\caption{Illustration of the vanilla U-Net we used. The number of feature channels is denoted on the top of the box. The arrows denote different operations. }
\label{fig:arch}
\end{figure}

\subsection{Network Architecture}
The architecture is illustrated in Figure \ref{fig:arch}. We adopt the vanilla U-Net, which has a common encoder-decoder architecture. We simply concatenate the PET/CT scan pairs as the input. For feature extraction, the vanilla U-Net transfer the input into feature maps with 32 channels at the first layer and doubles the number of feature channels at each down-sampling step. 
At each stage of the decoder, two feature maps are concatenated as the input, one from the skip connections, and the other from the upsampled feature map of the previous stage.

\begin{table}[!htp]
    \centering
    \caption{Some detailed settings of the data augmentations}
    \label{tab:DA_setting}
    \begin{tabular}{|l|l|}
    \hline
        Data augmentation & Parameters \\ \hline \hline
        Random Brightness Transform & $range=(0.75, 1.25), \sigma=0.10$  \\ \hline
        Random Gamma Transform &  $range=(0.70, 1.50), p=0.30$  \\ \hline
        Random Flip & $p=0.5$  \\ \hline
        Random Rotation & $range=(-15, +15)$  \\ \hline
    \end{tabular}
\end{table}

\subsection{Data Preprocessing}
According to the official pre-processing pipeline of autoPET\footnote{https://autoPET.grand-challenge.org/Dataset/}, the provided CT and PET scans are resampled to 1.500 $\times$ 1.0182 $\times$ 1.0182 mm. In addition, the PET scans are normalized and converted to Standardized Update Values~(SUV). We follow this pipeline and develop our method based on the provided pre-processed NifTI files.

\subsection{Data Augmentation}
\subsubsection{Random Brightness Transform}
Brightness transform contains multiplicative and additive operations on the intensity, formulated as
\begin{equation}
    X_{out} = m \times (X_{in}) + \mathcal{N}(0, \sigma),
\end{equation}
where $m$ is a scale factor randomly sampled from a uniform distribution of a fixed range, with the range shown in Table \ref{tab:DA_setting}.

\subsubsection{Random Gamma Transform}
Gamma transform is a technique used to compensate for the non-linear display characteristics of a device, formulated as
\begin{equation}
    X_{out} = (X_{in})^{1/\gamma}.
\end{equation}
The original image is corrected by sending the signal through the inverse function $f(x) = x^{1/\gamma}$ based on a fixed gamma value. The gamma value is randomly sampled from a uniform distribution of a fixed range. 

Since the CT and PET scans share similar anatomical structures, we use the same random flip and rotation as the spatial data augmentations. Referring to nnU-Net~\cite{isensee2021nnu}, the parameters of data augmentations are listed in Table~\ref{tab:DA_setting}.

\begin{table}[!htp]
    \centering
    \caption{Implementation details in the training stage of our final submission.}
    \label{tab:implementation_details_training}
    \begin{tabular}{|l|c|}
    \hline
        Training setup & Details \\ \hline \hline
        Spacing & [1.5, 1.01821005, 1.01821005] \\ \hline
        Crop size & $192\times192\times192$ \\\hline
        Batch size & 2 \\\hline
        Data augmentation & ``default\_3D\_augmentation\_params'' in nnUNet \\\hline
        Learning rate & 0.0001 \\\hline
        Optimizer & SGD + Poly decay policy \\\hline
        Number of iterations & 250,000 \\ \hline
    \end{tabular}
\end{table}

\begin{table}[!htp]
    \centering
    \caption{Implementation details in the testing stage of our final submission.}
    \label{tab:implementation_details_testing}
    \begin{tabular}{|l|c|}
    \hline
        Testing setup & Details \\ \hline \hline
        Spacing & [1.5, 1.01821005, 1.01821005] \\ \hline
        Crop size & $192\times192\times192$ \\ \hline
        Inference strategy & Sliding window strategy with step size 0.5 \\ \hline
        Postprocessing & None \\ \hline
        Ensemble strategy & Ensemble the results of 5 folds \\ \hline
    \end{tabular}
\end{table}

\section{Experiments}

In the training stage, we keep the original spacing, and the crop size is $192 \times 192 \times 192$. The batch size is fixed to 2 and each model is trained for 250,000 iterations. We use SGD optimizer with momentum of 0.99, weight decay of 0.001. The learning rate is set as 0.0001 at the beginning and decayed based on the poly decay policy $(1-iter/250000)^{0.9}$ for every 250 iterations. The final setups of our training and testing stage are listed in Table~\ref{tab:implementation_details_training}.

\subsection{Implementation Details}
Our development environment is CUDA 11.2, Pytorch 1.11.0, and Tesla A100 80G. We implement our method based on the nnU-Net codebase~\cite{isensee2021nnu}. 

In the predicting stage, we just follow the nnU-Net with no post-processing. Specifically, nnU-Net adopts the sliding-window strategy for inference due to the high resolution of volumetric medical images. To find the trade-off between accuracy and speed, we ablate the step size of the sliding-window strategy in the ablation section. The final setups of our testing stage are listed in Table~\ref{tab:implementation_details_testing}.

\begin{table}[!htp]
    \centering
    \caption{Results of different network architectures on the preliminary test set. {*} indicates UNETR is implemented by ourselves.}
    \label{tab:res_arch}
    \begin{tabular}{|c|c|c|c|c|c|c|}
    \hline 	 	  	 	  	 	 	  	
        Network architecture & stage  & feature channels                  & params & DSC  & FPV & FNV \\ \hline \hline  	 	
        UNETR{*}              & -      & -                                & 133.0M & 0.51 & 6.28 & \textbf{1.58} \\ \hline
        Shallow U-Net        & 4      & $(16, 32, 64, 128)$               & 1.4M & 0.71 & \textbf{1.22} & 3.39  \\ \hline
        Vanilla U-Net        & 5      & $(32, 64, 128, 256, 512)$         & 22.6M & \textbf{0.92} & 1.99 & 1.68  \\ \hline
        Deeper U-Net         & 6      & $(32, 64, 128, 256, 512, 1024)$  & 58.6M & 0.52 & 1.57 & 1.95  \\ \hline
    \end{tabular}
\end{table}

\subsection{Ablation}

\textbf{Different architectures.} As shown in Table~\ref{tab:res_arch}, we first evaluate the performance of different number of stages and feature channels in U-Net. The vanilla U-Net outperforms other architectures for DSC and FPV on the preliminary test set. Moreover, the CNN-based vanilla U-Net improves the transformer-based UNETR~\cite{hatamizadeh2022unetr} by 42\% in DSC. In addition, the params of the vanilla U-Net is 22.6M, which is much less than Deeper U-Net (58.6M) and UNETR (133.0M). Note that the parameters are concluded by torchinfo\footnote{https://github.com/TylerYep/torchinfo}.

\begin{table}[!htp]
    \centering
    \caption{DSC of different input shapes in the training stage on our validation set. ``Ori spacing'' indicates that we keep original spacing for all data in training. ``1/2 spacing'' indicates that we decrease the spacing to 1/2 for training cases.}
    \label{tab:res_crop_size_tr}
    \setlength{\tabcolsep}{7mm}
    \begin{tabular}{|c|c|c|}
    \hline 	 	  	 	  	 	 	  	
        \multirow{2}{*}{Input shape}       & \multicolumn{2}{c|}{DSC} \\
        \cline{2-3}
                                           & Ori spacing & 1/2 spacing \\ \hline \hline
        $128 \times 128 \times 128$        & 0.62 & 0.59 \\ \hline
        $160 \times 160 \times 160$        & 0.69 & 0.65 \\ \hline
        $192 \times 192 \times 192$        & \textbf{0.75} & 0.68\\ \hline
        $224 \times 224 \times 224$        & 0.74 & \textbf{0.72}\\ \hline
    \end{tabular}
\end{table}

\noindent\textbf{Input shape and spacing in the training stage.} We further ablate different crop sizes in the training stage, and show the DSC results in Table~\ref{tab:res_crop_size_tr}. For the original spacing, the cropped input size of $192\times192\times192$ achieves the best DSC 75\%, while the crop size of $224\times224\times224$ performs better with 1/2 spacing. Table~\ref{tab:res_crop_size_tr} also shows that keeping original spacing is the best way to train a better model.

\begin{table}[!ht]
\centering
\makebox[0pt][c]{\parbox{1\textwidth}{%
\begin{minipage}[b]{0.46\hsize}\centering
\caption{Results of different input shapes in the inference stage on our validation set. The inference time shows the total time of 103 cases.}
\label{tab:res_crop_size_ts}
\resizebox{\linewidth}{!}{
    \begin{tabular}{|c|c|c|c|c|c|c|c|}
    \hline 	 	  	 	  	 	 	  	
        Input shape                        & DSC  & FPV & FNV   &Inference time~(s) \\ \hline \hline
        $128 \times 128 \times 128$        & 0.73 & \textbf{2.43} & 7.88 & 5528   \\ \hline
        $160 \times 160 \times 160$        & \textbf{0.75} & 3.11 & 6.83 & 4248    \\ \hline
        $192 \times 192 \times 192$        & \textbf{0.75} & 4.24 & 6.10 & 4672    \\ \hline
        $224 \times 224 \times 224$        & \textbf{0.75} & 3.73 & 6.07 & 3878    \\ \hline
        $256 \times 256 \times 256$        & \textbf{0.75} & 4.65 & 5.73 & 4757    \\ \hline
        $288 \times 288 \times 288$        & 0.74 & 6.73 & 5.44 & \textbf{3288}    \\ \hline
        $320 \times 320 \times 320$        & 0.72 & 10.87 & \textbf{5.12} & 4169    \\ \hline
    \end{tabular}
}
\end{minipage}
\hfill
\begin{minipage}[b]{0.53\hsize}\centering
\caption{Results of different step sizes of the sliding window strategy in the inference stage on our validation set. The inference time shows the total time of 103 cases.}
\label{tab:res_step_size}
\resizebox{\linewidth}{!}{
    \begin{tabular}{|c|c|c|c|c|c|c|}
    \hline 	 	  	 	  	 	 	  	
        Step size                        & DSC  & FPV & FNV &Inference time~(s)  \\ \hline \hline
        $0.5 \times input\_shape$        & \textbf{0.75} & 4.24 & \textbf{6.10} & 4672 \\ \hline
        $0.6 \times input\_shape$        & \textbf{0.75} & 4.03 & 6.43 & 3166 \\ \hline
        $0.7 \times input\_shape$        & \textbf{0.75} & 3.67 & 6.43 & 2961 \\ \hline
        $0.8 \times input\_shape$        & \textbf{0.75} & 3.71 & 6.51 & 2933 \\ \hline
        $0.9 \times input\_shape$        & \textbf{0.75} & 3.70 & 6.51 & 2937 \\ \hline
        $1.0 \times input\_shape$        & \textbf{0.75} & \textbf{3.66} & 6.51 & \textbf{2915} \\ \hline
    \end{tabular}
}
\end{minipage}
}}
\end{table}

\noindent\textbf{Input shape and step size in the testing stage.} To find a faster and more accurate way of inference, we ablate different input sizes and the step sizes of the sliding window strategy. Table~\ref{tab:res_crop_size_ts} shows the results of different input shapes in the testing stage. The DSC changes little with different input sizes, while FNV and FPV are very sensitive. The smaller~(larger) input size we use, the lower FPV (FNV) we get. For step sizes, the results are shown in Table~\ref{tab:res_step_size}. The DSC is stable at 75\% in all step sizes. However, FNV and FPV are also sensitive to different step size. 

\begin{table}[!ht]
    \centering
    \caption{Final submissions. ``Status'' indicates whether our submission succeeded in the submission system.}
    \setlength{\tabcolsep}{7mm}
    \label{tab:final_submission}
    \begin{tabular}{|c|c|c|}
    \hline 	 	  	 	  	 	 	  	
        Step size & Crop size                   & Status  \\ \hline \hline
        0.5       & $192 \times 192 \times 192$ & Failed \\ \hline
        0.5       & $224 \times 224 \times 224$ & Failed \\ \hline  
        0.6       & $192 \times 192 \times 192$ & Success \\ \hline
        0.7       & $192 \times 192 \times 192$ & Success \\ \hline
    \end{tabular}
\end{table}


\begin{table}[!ht]
\centering
\makebox[0pt][c]{\parbox{1\textwidth}{%
\begin{minipage}[b]{0.48\hsize}\centering
\caption{Comparison to other teams on the preliminary test set.}
\label{tab:preliminary_results}
\setlength{\tabcolsep}{2mm}
\begin{tabular}{|l|c|c|c|}
\hline 	 	  	 	  	 	 	  	
    Teamname  & DSC  & FNV  & FPV  \\ \hline \hline
    Blackbean & \textbf{0.93} & 0.79 & 1.07 \\ \hline
    SM        & 0.93 & 1.42 & 0.90 \\ \hline
    BDAV      & 0.93 & 1.57 & \textbf{0.78} \\ \hline
    PET\_TEAM & 0.92 & \textbf{0.53} & 0.89 \\ \hline
    UIH-FL    & 0.93 & 1.57 & 0.81 \\ \hline
\end{tabular}
\end{minipage}
\hfill
\begin{minipage}[b]{0.48\hsize}\centering
\caption{Comparison to other teams on the final test set.}
\label{tab:final_results}
\setlength{\tabcolsep}{2mm}
\begin{tabular}{|l|c|c|c|}
\hline 	 	  	 	  	 	 	  	
    Teamname   & DSC  & FNV  & FPV  \\ \hline \hline
    Blackbean  & \textbf{0.62} & 0.54 & \textbf{2.84} \\ \hline
    BDAV       & 0.62 & 0.75 & 3.61 \\ \hline
    FightTumor & 0.60 & \textbf{0.47} & 5.10 \\ \hline
    UIH-FL     & 0.61 & 0.83 & 4.85 \\ \hline
    heiligerl  & 0.61 & 0.63 & 5.87 \\ \hline
\end{tabular}
\end{minipage}
}}
\end{table}

\subsection{Final submissions}
According to the ablation studies, we adopt the original spacing and set the input shape to $192\times192\times192$ in the training stage. We simply train 5 folds models using the same splits with nnUNet, and combine the results of all models in the inference stage (Also the same as nnUNet). We first evaluate our model with step size 0.5 and crop size $192 \times 192 \times 192$ on the preliminary test set, the results are shown in Table~\ref{tab:preliminary_results}, which achieves the Mean Position 17.3 (Can be found in the autoPET homepage) and ranks top-1 on the leaderboard (Team name: ``Blackbean"). In the final submissions, to take both accuracy and inference time into consideration, we submitted different strategies, which are listed in Table~\ref{tab:final_submission}. In the final leaderboard, our vanilla U-Net also achieves first place among all competitors, and the results are shown in Table~\ref{tab:final_results}.






\bibliographystyle{splncs04}
\bibliography{ref}
%




\end{document}